# Classical Model for Broadband Squeezed Vacuum Driving Two-Photon Absorption or Sum Frequency Generation

*Michael G. Raymer[1,2] * and Tiemo Landes[1,2]*

[1] *Department of Physics, University of Oregon, Eugene, OR 97403, USA*
[2] *Oregon Center for Optical, Molecular and Quantum Science, University of Oregon, Eugene, OR 97403, USA*
* raymer@uoregon.edu

**Abstract:** We address theoretically the question of classical stochastic fields mimicking quantum states of light in the context of nonlinear spectroscopy and nonlinear optics, in particular two-photon absorption (TPA) and sum-frequency generation (SFG) driven by weak or bright broadband squeezed vacuum with time-frequency entanglement between photons. Upon using a well-defined but ad hoc subtraction of vacuum-energy terms ("renormalization"), we find that the classical stochastic model yields exactly the same predictions as the full quantum-field theory for all of the phenomena considered here, in both the low-gain and high-gain regimes of squeezed vacuum. Such predictions include the linear-flux scaling of TPA and SFG rates at low incident photon flux, as well as the dependence of TPA and SHG rates on the relative linewidths of the squeezed light and the ground-to-final-state transition in the material system, and the spectrum of SFG generated by bright squeezed vacuum.

**Note:** Based on results reported in a talk by the authors, "Where is the Quantum Advantage of Time-Frequency-Entangled Photon Pairs for Nonlinear Spectroscopy?" Physics Colloquium, City University of New York, December 8, 2021.

## 1. Introduction

We present a theoretical model wherein stochastic classical fields mimic quantum fields in the context of a broadband (multi-spectral mode) squeezed vacuum field driving non-resonant molecular two-photon absorption (TPA) or sum-frequency generation (SFG). We model the squeezed-vacuum state of light as a classical Gaussian random process, representing a stochastic 'zero-point field' that is amplified in a phase-sensitive parametric (difference-frequency) amplifier, creating correlations in pairs of frequencies—thus mimicking a key feature of quantum squeezed light.

The present work uses the results presented in [ray22], where we presented an analytically tractable, quantum model for low-gain squeezed vacuum (or spontaneous parametric down conversion, SPDC) and high-gain squeezed vacuum (bright squeezed vacuum, BSV) and applied it to TPA. [ray22] The classical-optics model of squeezed vacuum is constructed by replacing quantum operators defined by their commutator relations by classical stochastic variables defined by Gaussian statistical properties and their correlation functions.

An analogous treatment was presented in [boi13], only in the high-gain limit, where it was found that TPA by squeezed-vacuum states could be modeled well using a stochastic classical model. the present challenge is to extend that conclusion to the low-gain spontaneous parametric down conversion (SPDC) regime. A related approach was reported in [kul23] for characterizing the spatiotemporal correlations of BSV.

We find that, after carrying out a well-defined but ad hoc subtraction of vacuum-energy terms (which we call "renormalization"), such a stochastic classical model can reproduce all the predictions of quantum optics theory considered here, in both the low-gain and high-gain regimes of squeezed vacuum. Such predictions include the well-known linear-flux scaling of TPA and SFG rates at low incident photon flux, as well as the dependence of TPA and SHG rates on the relative linewidths of the squeezed light and the ground-to-final-state transition in the material system, and the spectrum of SFG generated by BSV. All of these depend explicitly on the form of the four-frequency correlation function, which is shown to be identical in the quantum and stochastic classical models in the high-gain regime and in the low-gain regime following the renormalization procedure. For reviews and derivations, see [ray 21, lan21].

## 2. Quantum and classical models for propagating fields

We begin with the full quantum theory for generation of squeezed vacuum states and then 'de-quantize' it. For a collimated light beam in the absence of nonlinear interactions, the electric field operator defined within a given frequency band $B_J$ (a spectral region with range significantly smaller than the carrier frequency with center frequency $\omega_J$) is well approximated as [ray22], with positive frequencies only in the integral,

$$\hat{\mathbf{E}}^{(+)}(\mathbf{r},t) = i\sum_m \int_{B_J} \dbar\omega \sqrt{\frac{\hbar\omega}{2\varepsilon_0 c n_m(\omega)}}\ \hat{a}_m(\omega) e^{-i\omega t} \exp(ik_m(\omega)z) \mathbf{e}_m \mathrm{w}_m(\mathbf{r}), \qquad (1)$$

where $m$ labels different polarization ($\mathbf{e}_m$) and normalized transverse spatial modes ($\mathrm{w}_m(\mathbf{r})$), and the propagation constant in free space equals $k_m(\omega) = \omega n_m(\omega)/c$, with $n_m(\omega)$ being the refractive index for mode $m$ and $c$ being the vacuum speed of light. Carets indicate operators and for convenience we denote $\dbar\omega = d\omega/2\pi$. The creation and annihilation operators satisfy the commutator $[\hat{a}_m(\omega), \hat{a}_n^\dagger(\omega')] = 2\pi\delta(\omega-\omega')\delta_{mn}$. The expectation value of the commutator for the vacuum state is

$$\langle vac | [\hat{a}_m(\omega), \hat{a}_n^\dagger(\omega')] | vac \rangle = 2\pi\delta(\omega-\omega')\delta_{mn}. \qquad (2)$$

To 'de-quantize' the field operator and produce a classical stochastic-field (SF) model, we replace the creation and annihilation operators by a set of *classical* complex Gaussian random processes $a_m(\omega)$,

$$\begin{aligned}\hat{a}_m(\omega) &\to a_m(\omega) \\ \hat{a}_m^\dagger(\omega) &\to a_m^*(\omega)\end{aligned} \qquad (3)$$

(absence of carets indicates classical quantities). The stochastic field is

$$\mathbf{E}^{(+)}(\mathbf{r},t) = i\sum_m \int_{B_J} \dbar\omega \sqrt{\frac{\hbar\omega}{2\varepsilon_0 c n_m(\omega)}}\ a_m(\omega) e^{-i\omega t} \exp(ik_m(\omega)z) \mathbf{e}_m \mathrm{w}_m(\mathbf{r}). \qquad (4)$$

The defining relations for the random processes are the second-order correlation functions,

$$\langle a_m(\omega) a_n^*(\omega') \rangle_{SF} = P_{SF}(\omega) 2\pi \delta(\omega - \omega') \delta_{mn}$$
$$\langle a_m(\omega) a_n(\omega') \rangle_{SF} = 0$$

(5)

where the brackets denote ensemble averages over realizations of the random processes. The frequency delta function is consistent with the random process being stationary (statistically invariant under a time shift). $P_{SF}(\omega)$ is a unitless spectral density of fluctuations, which we are free to choose as we build the classical model. The Kronecker delta states that the random processes for different transverse-spatial or polarization modes are statistically independent. Along with the assumption of being complex Gaussian random processes, the second-order correlation functions fully define the statistical 'state' of the classical field.

We refer to the ever-present random processes as zero-point vacuum fields, but analogy to quantum theory. Their time-domain representations are stationary Gaussian random processes given by the Fourier transform,

$$\tilde{a}_m(t) = \int d\omega\, e^{-i\omega t} a_m(\omega) .$$

(6)

In the limiting case that the spectrum $P_{SF}(\omega)$ is independent of frequency (white noise) we have $\langle \tilde{a}_m(t) \tilde{a}_m^*(t') \rangle = P_{SF} \delta(t - t')$.

We treat a simplified model, with a single polarization and transverse mode, and write for the quantum and classical fields,

$$\hat{E}^{(+)}(z,t) = iL_0 \int d\omega\, \hat{a}(\omega) \exp(ik(\omega)z) e^{-i\omega t}$$
$$E^{(+)}(z,t) = iL_0 \int d\omega\, a(\omega) \exp(ik(\omega)z) e^{-i\omega t}$$

(7)

where $L_0 = (\hbar \omega_0 / 2c\varepsilon_0 n_0 A_0)^{1/2}$, $\omega_0$ is the center frequency of the spectral band of interest and $n_0$ is its nominal refractive index. The effective beam area $A_0$ relevant to a molecule located at a fixed point $\mathbf{r}_0$ is given by the mode amplitude, $(1/A_0)^{1/2} \equiv \mathrm{w}(\mathbf{r}_0)$. In the quantum context, this form implies that the photons are indistinguishable except for their frequencies. The distinguishable case is easily treated by retaining the sum over modes.

It is often useful to use the concept of temporal modes, which are temporally (and spectrally) orthogonal traveling wave packets described by a set of orthonormal functions $\tilde{\psi}_j(t)$. [ray22, ray23] In the case that the stochastic field is white noise, the energy in any of these wave-packet modes within a spectral band centered at $\omega_0$ is taken to equal $P_{SF} \hbar \omega_0$, where $P_{SF}$ is a dimensionless adjustable parameter of the classical model. See Appendix A. The appearance of Plank's constant allows for easy comparison of the SF model results and the quantum theory (QT) results.

When choosing the vacuum spectral density to equal $P_{SF} = 1/2$, our model becomes close to a model called *stochastic electrodynamics*, which has been applied to classical derivations of the blackbody spectrum and Casimir and van der Waals forces. [boy19] This assignment ascribes a classical zero-point energy per frequency mode of $\hbar\omega/2$.

## 3. Quantum theory of two-photon absorption and sum-frequency generation

Nonresonant TPA occurs when the sum of two incident photon frequencies is on or near resonance with the molecule's TPA transition, but the individual fields are far from any intermediate molecular resonance. Then the dominant term (so-called DQC term) in a perturbation expansion represents direct excitation to an excited electronic state by simultaneous absorption of two photons without creating 'real' population in any intermediate state.

Sum-frequency generation (SFG) occurs when two photons with frequencies $\omega, \tilde{\omega}$ are annihilated in a nonlinear-optical medium and a photon is created having frequency $\omega_3 = \omega + \tilde{\omega}$. Dayan showed that SFG involving broadband quantum fields is well modeled by equations analogous to those of TPA. [day07]

We summarize TPA (or SFG) by writing the probability to find the molecule in the excited state (or the probability to create an SFG photon) after a pulse of light has passed through the medium. Both are given by a common expression, see [ray21] and Appendix B,

$$P = \int d\omega' \int d\omega \int d\tilde{\omega}\, K(\omega',\omega,\tilde{\omega}) C^{(4)}(\omega', \omega+\tilde{\omega}-\omega', \omega, \tilde{\omega}), \qquad (8)$$

where $K(\omega',\omega,\tilde{\omega})$ is a relevant medium response function and the four-frequency correlation function for an arbitrary state of light is

$$C^{(4)}(\omega_a,\omega_b,\omega_c,\omega_d)_{QT} = \left\langle \hat{c}^\dagger(\omega_a)\hat{c}^\dagger(\omega_b)\hat{c}(\omega_c)\hat{c}(\omega_d) \right\rangle. \qquad (9)$$

where the annihilation operator $\hat{c}(\omega)$ plays the same role in the field incident on the medium as the operator $\hat{a}(\omega)$ plays in Eq.(7). In quantum theory (QT) the brackets indicate an expectation value (trace with the field-state density operator). In the SF model, we replace the operator trace by an ensemble average over complex Gaussian random processes $c(\omega)$,

$$C^{(4)}(\omega_a,\omega_b,\omega_c,\omega_d)_{SF} = \left\langle c^*(\omega_a)c^*(\omega_b)c(\omega_c)c(\omega_d) \right\rangle \qquad (10)$$

Here we develop this idea using our model for squeezed states that is unified across the low- and high-gain regimes.

Therefore, to explore whether the classical SF model can reproduce the QT results, it is sufficient to learn under what conditions and assumptions the four-frequency correlation function of the light incident on the medium is the same under the two models. We do this next, where we can construct a precise correspondence between classical and quantum theory in the high-gain BSV regime, as well as in the low-gain SPDC regime after using an introduced renormalization procedure.

**4. Broadband squeezed field in quantum and classical stochastic models**

We begin by briefly reviewing the quantum formulation, following [ray22], which has similarities to [boit13] in the high-gain limit. We focus on Type-0 or Type-I phase matching for generation of co-polarized, co-propagating photon pairs, in which case photons have no distinguishing labels other than frequency.

Entangled photon pairs and the accompanying quantum-noise squeezing can be generated by optical parametric amplification (OPA) seeded by spontaneous parametric down conversion (SPDC). In our model the parametric down conversion (PDC) crystal with length $z$ and second-order nonlinear coefficient $\chi^{(2)}$ is pumped by a strong, coherent monochromatic continuous-wave (CW) field with constant amplitude $E_p$ and angular frequency $\omega_p$. (Pulsed squeezed light is treated below.) In quantum theory this interaction causes a Heisenberg-picture transformation of the input (vacuum) field operators $\hat{a}(\omega)$ to output field operators $\hat{b}(\omega)$, given by the frequency-dependent two-mode squeezing transformation [ray22],

$$\hat{b}(\omega) = f(\omega)\hat{a}(\omega) + g(\omega)\hat{a}^\dagger(2\omega_0 - \omega) \quad (11)$$

where $f$ and $g$ are given by

$$f(\omega) = \cosh[s(\omega)z] - i\frac{\Delta k(\omega)}{2s(\omega)}\sinh[s(\omega)z]$$
$$g(\omega) = i\frac{\gamma}{s(\omega)}\sinh[s(\omega)z] \quad (12)$$

with the phase mismatch being approximated for type-0 or type-1 as $\Delta k(\omega) = -\kappa(\omega - \omega_0)^2$ and

$$s(\omega) = \sqrt{\gamma^2 - \Delta k(\omega)^2} \quad (13)$$

and $\omega_0 = \omega_p/2$. We used the abbreviation $\kappa \equiv k''/2$, with $k''$ being the group-delay dispersion in the crystal at frequency $\omega_0$. The (real) gain coefficient is $\gamma = (\omega_0/c)\chi^{(2)}E_p$. The gain functions have the symmetry $f(2\omega_0 - \omega) = f(\omega)$ and $g(2\omega_0 - \omega) = g(\omega)$ and satisfy the Bogoliubov unitarity condition

$$|f(\omega)|^2 = |g(\omega)|^2 + 1 \quad (14)$$

The initial field state is the vacuum, which becomes amplified in a phase-sensitive manner. In quantum theory the amplification generates a quadrature-squeezed-vacuum state, creating correlated pairs of photons as well as two-mode squeezing correlations among pairs of frequencies offset symmetrically from the central frequency $\omega_0$. The amplified field is found to have a spectral photon rate $S_{QT}(\omega)$ (photons per second per frequency interval) related to the two-frequency correlation function by

$$\langle \hat{b}^\dagger(\omega)\hat{b}(\tilde{\omega})\rangle_{QT} = \langle vac|\hat{b}^\dagger(\omega)\hat{b}(\tilde{\omega})|vac\rangle$$
$$= S_{QT}(\omega)2\pi\delta(\omega-\tilde{\omega}) \tag{15}$$

with the spectrum given by

$$S_{QT}(\omega) = |g(\omega)|^2 = \frac{\gamma^2}{|s(\omega)|^2}\sinh^2[s(\omega)z] \tag{16}$$

and the delta function being consistent with the stationarity of the field.

The forms of $f(\omega)$ and $g(\omega)$ in the low-gain limit are

$$f(\omega) \to \exp[i\kappa(\omega-\omega_0)^2 z] \to 1$$
$$g(\omega) \to i\gamma z \frac{\sin[\kappa(\omega-\omega_0)^2 z]}{\kappa(\omega-\omega_0)^2 z} \tag{17}$$

where here we replaced $f(\omega)$ by 1 under the assumption that a dispersion-correcting optical device (e.g. a pulse compressor) is inserted in the beam path, a procedure justified in detail in [ray22]. In the high-gain limit we then have,

$$f(\omega) \to \frac{1}{2}\exp[\gamma z]\exp\left[-\left(\frac{\kappa^2 z}{2\gamma}\right)(\omega-\omega_0)^4\right]$$
$$g(\omega) \to if(\omega) \tag{18}$$

To implement the classical SF model, we write by analogy with Eq.(11) the amplified stochastic field as

$$b(\omega) = f(\omega)a(\omega) + g(\omega)a^*(2\omega_0 - \omega) . \tag{19}$$

The classical squeezed field is then

$$E^{(+)}(z,t) = L_0\int d\omega\, b(\omega)e^{-i\omega t}$$
$$= L_0\int d\omega\, f(\omega)a(\omega)e^{-i\omega t} + L_0\int d\omega\, g(2\omega_0-\omega)a^*(\omega)e^{-i(2\omega_0-\omega)t} \tag{20}$$

showing the frequency anticorrelation expected for classically squeezed light.

In the SF model, the stationary squeezed field has a spectral flux $S_{SF}(\omega)$ related to the two-frequency correlation function by

$$\langle b^*(\omega)b(\tilde{\omega})\rangle = \left(|f(\omega)|^2 + |g(\omega)|^2\right) P_{SF}(\omega) 2\pi\delta(\omega - \tilde{\omega}) \qquad (21)$$

So, in this model its spectrum is

$$\begin{aligned} S_{SF}(\omega) &= P_{SF}(\omega)\left(|f(\omega)|^2 + |g(\omega)|^2\right) \\ &= P_{SF}(\omega)\left(2|g(\omega)|^2 + 1\right) \end{aligned}, \qquad (22)$$

where we used Eq.(14).

Note that if we take $P_{SF}(\omega)$ to be independent of frequency, that is $P_{SF} = 1/2$ as discussed above, we have a spectrum whose integral over frequency diverges. This point illustrates the difficulty of adopting the stochastic field model as a realistic, physical model. Thus, in the following we will consider an ad hoc model that removes the infinity (renormalization). Before we consider this approach, we will show how to convert the steady-state squeezing theory to a pulsed squeezed-light model by use of temporal gating.

## 5. Temporally gated fields

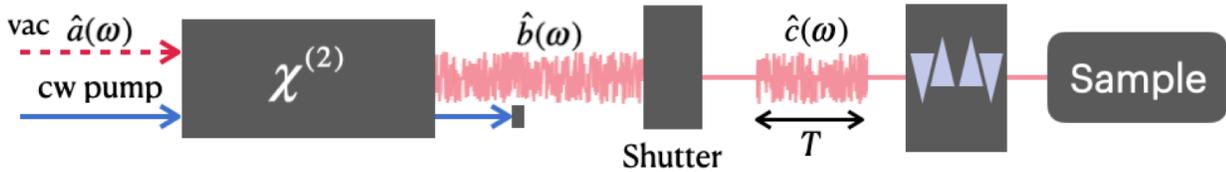

Fig. 1. The model for long-pulse squeezed vacuum light. The initial vacuum field is amplified in a second-order nonlinear optical crystal pumped by a CW laser and phase-matched for degenerate type-0 or type-1 colinear spontaneous parametric downconversion (PDC). The PDC passes through a temporal shutter open for duration T, then passes through a dispersion-compensating optic, and into a two-photon-absorbing molecular sample or SFG medium.

As illustrated in **Fig.1** and detailed in [ray22], a way to model the pulsed nature of a squeezed field (or a finite interaction time of a molecule with a CW field), while using the steady-state solutions, is by assuming the CW field passes through a temporal gate that multiplies the field by a time-windowing function $\tilde{W}(t)$ that equals 1 inside the time window $\{-T/2, T/2\}$ and zero otherwise. The Fourier transform of this function is $W(\omega) = T \, \text{sinc}[\omega T/2]$. For simplicity, we assume that T is much greater than the light's coherence time (inverse of its spectral width). The resulting time-gated operator $\hat{c}(\omega)$ is, in the frequency domain,

$$\hat{c}(\omega) = f(\omega)\hat{A}(\omega) + g(\omega)\hat{A}^\dagger(2\omega_0 - \omega) \qquad (23)$$

where, under the assumption that $T$ is much greater than the light's coherence time, the filtered input operator is well approximated by the convolution

$$\hat{A}(\omega) = \int d\omega' W(\omega - \omega')\hat{a}(\omega') \qquad (24)$$

The commutators for the filtered operators are denoted by $D(\omega - \tilde{\omega})$ and are given by

$$\begin{aligned}[\hat{A}(\omega), \hat{A}^\dagger(\tilde{\omega})] &= [\hat{c}(\omega), \hat{c}^\dagger(\tilde{\omega})] \\ &= \int d\omega' W(\omega - \omega')W(\tilde{\omega} - \omega') \quad \text{check} \\ &= T \, \text{sinc}\left[(\omega - \tilde{\omega})T/2\right] \\ &\equiv D(\omega - \tilde{\omega})\end{aligned} \qquad (25)$$

The commutators are broadened in frequency because we discarded operators for light that is blocked by the gate, so the transformation is not unitary (valid for the calculations being considered). The $D$ function is normalized as

$$\begin{aligned} D(0) &= T \\ \int d\omega D(\omega - \tilde{\omega}) &= 1 \\ \int d\omega D(\omega - \tilde{\omega})^2 &= T \end{aligned}, \qquad (26)$$

In **quantum theory**, the two-frequency correlation function of the field is

$$\begin{aligned}\langle \hat{c}^\dagger(\omega)\hat{c}(\tilde{\omega})\rangle_{QT} &= \langle vac|\hat{c}^\dagger(\omega)\hat{c}(\tilde{\omega})|vac\rangle = \\ &= \langle vac|\left(\cancel{f(\omega)\hat{A}^\dagger(\omega)} + g(\omega)\hat{A}(2\omega_0 - \omega)\right)\left(\cancel{f(\tilde{\omega})\hat{A}(\tilde{\omega})} + g(\tilde{\omega})\hat{A}^\dagger(2\omega_0 - \tilde{\omega})\right)|vac\rangle \\ &= g(\omega)g(\tilde{\omega})\langle vac|\hat{A}(2\omega_0 - \omega)\hat{A}^\dagger(2\omega_0 - \tilde{\omega})|vac\rangle \\ &= g(\omega)g(\tilde{\omega})D(\omega - \tilde{\omega})\end{aligned} \qquad (27)$$

The mean number of photons, which are created in pairs, in the time-gated squeezed pulse of duration $T$ thus equals

$$\begin{aligned}N_{QT} &= \int d\omega \langle vac|\hat{c}^\dagger(\omega)\hat{c}(\omega)|vac\rangle \\ &= T\int d\omega |g(\omega)|^2\end{aligned}, \qquad (28)$$

where we assumed that the time window $T$ is long compared to other time scales so that $D$ acts like a 'fat' delta function with peak value $D(0) = T$.

In the **classical SF model**, we have for the gated field

$$c(\omega) = f(\omega)A(\omega) + g(\omega)A^*(2\omega_0 - \omega)$$
$$A(\omega) = \int d\omega' W(\omega - \omega')a(\omega') \tag{29}$$

with

$$\begin{aligned}
\langle A(\omega)A^*(\tilde{\omega})\rangle_{SF} &= \int d\omega' W(\omega-\omega')\int d\omega'' W(\tilde{\omega}-\omega'')\langle a(\omega')a^*(\omega'')\rangle_{SF} \\
&= \int d\omega' W(\omega-\omega')\int d\omega'' W(\tilde{\omega}-\omega'')P_{SF}(\omega')2\pi\delta(\omega'-\omega'') \\
&= \int d\omega' W(\omega-\omega')W(\tilde{\omega}-\omega')P_{SF}(\omega') \\
&\approx P_{SF}(\omega)\int d\omega' W(\omega-\omega')W(\tilde{\omega}-\omega') \\
&= P_{SF}(\omega)D(\omega-\tilde{\omega})
\end{aligned} \tag{30}$$

where again we assumed that the time window $T$ is long, so that $W(\omega)$ is a spectrally narrow function compared to $P_{SF}(\omega)$. Then the correlation function in the SF model is

$$\begin{aligned}
\langle c^*(\omega)c(\tilde{\omega})\rangle_{SF} &= \\
\langle (f^*(\omega)A^*(\omega))(f(\tilde{\omega})A(\tilde{\omega}))\rangle_{SF} &+ \langle (g^*(\omega)A(2\omega_0-\omega))(g(v)A^*(2\omega_0-\tilde{\omega}))\rangle_{SF} \\
&= P_{SF}(\omega)\left(f^*(\omega)F(\tilde{\omega})+g^*(\omega)G(\tilde{\omega})\right)D(\omega-\tilde{\omega}) \\
&\approx P_{SF}(\omega)\left(|f(\omega)|^2+|g(\omega)|^2\right)D(\omega-\tilde{\omega}) \\
&= P_{SF}(\omega)\left(2|g(\omega)|^2+1\right)D(\omega-\tilde{\omega})
\end{aligned} \tag{31}$$

where we again assumed that the time window $T$ is long compared to other time scales so that $D$ acts like a delta function. The classical energy in the time window, stated in terms of equivalent photon numbers, is then

$$\begin{aligned}
N_{SF} &= \int d\omega \langle c^*(\omega)c(\omega)\rangle_{SF} \\
&= T\int d\omega P_{SF}(\omega)\left(2|g(\omega)|^2+1\right)
\end{aligned}, \tag{32}$$

again showing the infinite contribution from the classical vacuum noise of zero-point field.

To remove thus unwanted contribution, we adopt an ad hoc renormalization method in which we subtract the result with g=0 from the result in Eq.(32),

$$\begin{aligned}
N_{SF} - N_{SF}|_{g=0} &= T\int d\omega 2P_{SF}(\omega)\left(|g(\omega)|^2+1/2\right) - T\int d\omega 2P_{SF}(\omega)(1/2) \\
&= T\int d\omega 2P_{SF}(\omega)|g(\omega)|^2
\end{aligned} \tag{33}$$

Then, taking $P_{SF}(\omega) = 1/2$, we find exact agreement with the QT result Eq.(28). A related renormalization method has been used in, for example, [ber10]. (Strictly speaking, renormalization should by applied at the level of the Hamiltonian underling the calculation; this point will be explored in a future study.)

**6. Four-frequency correlation function**

As mentioned above, all the features of TPA and SFG that we consider in this study depend only on the form of the four-frequency correlation function, which we show here to be identical in the quantum and stochastic classical models when considering either the high-gain limit without renormalization or for arbitrary gain when using the renormalization procedure.

In the ***quantum theory***, commutators are invoked to obtain the result, which can be written as a sum of 'coherent' and 'incoherent' terms, [ray22 has error in the distinguishable case, but that is no worry here and easily fixed] [ray22]

$$C^{(4)}(\omega_a, \omega_b, \omega_c, \omega_d)_{QT} = \langle \hat{c}^\dagger(\omega_a)\hat{c}^\dagger(\omega_b)\hat{c}(\omega_c)\hat{c}(\omega_d) \rangle \qquad (34)$$
$$= C_{coh} + C_{incoh}$$

where the two terms are called 'coherent' and 'incoherent' contributions, following terminology in [day07]

$$C_{coh} = g^*(\omega_a) f^*(\omega_a) f(\omega_c) g(\omega_c) D(2\omega_0 - \omega_a - \omega_b) D(2\omega_0 - \omega_d - \omega_c)$$
$$C_{incoh} = |g(\omega_a)|^2 |g(\omega_b)|^2 \left( D(\omega_b - \omega_c) D(\omega_a - \omega_d) + \xi D(\omega_a - \omega_c) D(\omega_b - \omega_d) \right) \qquad (35)$$

As shown in [ray22], the parameter $\xi$ equals 1 for the indistinguishable cases (colinear, co-polarized Type-0 or Type-I) and equals 0 for the distinguishable cases Type-II or off-axis Type-0 or Type-I phase matching.

The coherent contribution is highly efficient, as it contains the frequency anticorrelated part. The resulting enhancement of two-photon processes such as TPA and SFG has been studied extensively. [day07, lan21, ray22] The incoherent contribution contains the 'accidental' coincidences of uncorrelated photon frequencies, and so is less efficient.

For the ***classical SF model***, we use the complex Gaussian moment theorem [man95], stated as

$$\langle A^*(\omega_a) A^*(\omega_b) A(\omega_c) A(\omega_d) \rangle = \langle A^*(\omega_a) A(\omega_c) \rangle \langle A^*(\omega_b) A(\omega_d) \rangle + \qquad (36)$$
$$\langle A^*(\omega_a) A(\omega_d) \rangle \langle A^*(\omega_b) A(\omega_c) \rangle$$

to yield the four-frequency correlation function, using $f(2\omega_0 - \omega) = f(\omega)$ and $g(2\omega_0 - \omega) = g(\omega)$. Then we find, after lengthy algebra, a sum of 'correlated' and 'uncorrelated' terms,

$$C^{(4)}(\omega_a,\omega_b,\omega_c,\omega_d)_{SF} = \langle c^*(\omega_a)c^*(\omega_b)c(\omega_c)c(\omega_d)\rangle \qquad (37)$$
$$= C_{cor} + C_{uncor}$$

which, in the indistinguishable case yields

$$\begin{aligned}
C_{cor} &= 4P_{SF}^2 f^*(\omega_a)g^*(\omega_a)f(\omega_c)g(\omega_c) \times \\
&\quad D(2\omega_0 - \omega_a - \omega_b)D(2\omega_0 - \omega_c - \omega_d) \\
C_{uncor} &= 4P_{SF}^2 \left(|g(\omega_a)|^2 + 1/2\right)\left(|g(\omega_b)|^2 + 1/2\right) \times \\
&\quad \left(D(\omega_a - \omega_d)D(\omega_b - \omega_c) + D(\omega_a - \omega_c)D(\omega_b - \omega_d)\right)
\end{aligned} \qquad (38)$$

We use the names 'correlated' and 'uncorrelated' for the SF results to remind us that here we have classical random fields with distinct correlation properties, in contrast to the quantum theory where we have 'coherent' and 'incoherent' terms indicating quantum correlation or lack thereof.

First note that in the high-gain limit, Eq.(18) implies $f(\omega) \approx -ig(\omega)$, and we see that for $P_{SF} = 1/2$, this result is identical to the quantum one above in the high-gain limit. Second, note that Eq.(38) differs from the full quantum theory result, Eq.(35), only by the additive terms $+1/2$ and the absence of the multiplicative factor $\xi$ that corresponds to photon distinguishability or lack thereof.

Remarkably, if we adopt our renormalization method, subtracting the result in Eq.(38) with $g = 0$ from the result with $g \neq 0$, we see that, for $P_{SF} = 1/2$ and $\xi = 1$, the classical result is identical to the quantum one.

## 7. Summary

We have shown that the four-frequency correlation function is identical in the quantum and stochastic classical models in two cases: 1) in the high-gain regime and 2) for arbitrary gain when supplemented with the ad hoc renormalization procedure of subtracting the $g = 0$ results from the derived result. One can interpret this as consistent with the fact that the vacuum (zero-point) field cannot drive photon absorption processes; the only part of the field that can drive such processes are those that are excitations above the vacuum state, such as the squeezed vacuum state, weak or bright.

In this sense, the stochastic classical model can provide insight into the origins of the well-known linear-flux scaling of TPA and SFG rates at low incident photon flux, as well as the dependence of TPA and SHG rates on the relative linewidths of the squeezed light and the ground-to-final-state transition in the material system, and the spectrum of SFG generated by BSV. On the other hand, no classical model, such as the one presented here, can correctly describe truly quantum phenomena such as violation of Bell inequalities, quantum teleportation, etc.

**Acknowledgements**
We thank Gerd Leuchs for helpful discussions. This work was supported by the National Science Foundation RAISE-TAQS Program (PHY-1839216).

**Data Availability**
The data supporting this study are contained within the article.

**Disclosures**
The authors declare no conflicts of interest.

**Appendix A: Energy per temporal mode**

For nonlinear interactions involving propagating fields, we use the concept of temporal modes, which are temporally (and spectrally) orthogonal wave packets described by a set of orthonormal functions $\tilde{\psi}_j(t)$. [ray23] In the case that the stochastic field is white noise, the energy in any one of these wave-packet modes within a spectral band centered at $\omega_0$ can be calculated as

The energy in a given temporal mode is found by projecting the field onto it, and using Eq.(7) at $z = 0$,

$$E_j = \int dt\, E^{(+)}(0,t)\tilde{\psi}_j(t) \tag{39}$$

and, using $\psi_j(t) = \int d\omega\, e^{-i\omega t} \psi_j(\omega)$ to calculate the mean energy as follows,

$$E^{(+)}(z,t) = iL_0 \int \frac{d\omega}{2\pi} a(z,\omega)e^{-i\omega t},$$

$$\tilde{\psi}_j(\omega) = \int dt e^{-i\omega t}\psi_j(t), \quad \psi_j(t) = \int d\!\!\!\bar{}\,\omega' e^{-i\omega' t}\psi_j(\omega')$$

$$E_j = \int dt E^{(+)}(0,t)\psi_j(t) = L_0 \int dt \int \frac{d\omega}{2\pi} a(0,\omega)e^{-i\omega t}\int d\!\!\!\bar{}\,\omega' e^{-i\omega' t}\psi_j(\omega')$$

$$E_j = L_0 \int \frac{d\omega}{2\pi} a(\omega)\tilde{\psi}_j(\omega), \quad \varepsilon = \varepsilon_0 n_0^2$$

$$W = 2A_0 \frac{c}{n_0}\int_{A_0} d^2x \frac{n_0^2}{2}\varepsilon_0 \langle E_j E_j^*\rangle$$

$$= A_0 c n_0 \varepsilon_0 \int_{A_0} d^2x L_0^2 \left\langle \int \frac{d\omega}{2\pi} a(\omega)\psi_j(\omega)\cdot \int \frac{d\omega'}{2\pi} a^*(\omega')\psi_j^*(\omega')\right\rangle$$

$$= A_0 c n_0 \frac{\hbar\omega_0}{2c\varepsilon_0 n_0 A_0}\varepsilon_0 \int d\!\!\!\bar{}\,\omega\, \psi_j(\omega)\int d\!\!\!\bar{}\,\omega'\, \psi_j^*(\omega')\langle a(\omega)a^*(\omega')\rangle$$

$$= \frac{1}{2}\hbar\omega_0 \int d\!\!\!\bar{}\,\omega\, \psi_j(\omega)\int d\!\!\!\bar{}\,\omega'\, \psi_j^*(\omega') P_{SF} 2\pi\delta(\omega-\omega')$$

$$= P_{SF}\frac{1}{2}\hbar\omega_0 \int \frac{d\omega}{2\pi}\psi_j(\omega)\psi_j^*(\omega) \tag{40}$$

$$= P_{SF}\frac{1}{2}\hbar\omega_0 \to \hbar\omega_0/2$$

where we chose $P_{SF} = 1/2$ and used

$$\int d\!\!\!\bar{}\,\omega\, \psi_j(\omega)\psi_j^*(\omega) = 1 \tag{41}$$

## Appendix B: Sum-frequency generation

The Hamiltonian for SFG is

$$H = \xi'' \int d\!\!\!\bar{}\,\omega_1 \int d\!\!\!\bar{}\,\omega_2 \int d\!\!\!\bar{}\,\omega_3 \Phi(\omega_1,\omega_2,\omega_3)\hat{a}_1(\omega_1)\hat{a}_2(\omega_2)\hat{a}_3^\dagger(\omega_3)e^{-i(\omega_1+\omega_2-\omega_3)t} + hc \tag{42}$$

where $\xi''$ is proportional to the second-order susceptibility $\chi^{(2)}$ and the phase-matching function is

$$\Phi(\omega_1,\omega_2,\omega_3) = \text{sinc}\left(\frac{L}{2}[k_1(\omega_1)+k_2(\omega_1)-k_3(\omega_3)]\right) \tag{43}$$

A pulse-scattering interaction in lowest-order perturbation theory is described by the time-integrated Hamiltonian,

$$\int dt H = \xi' \int d\omega_1 \int d\omega_2 \Phi(\omega_1,\omega_2) \hat{a}(\omega_1)\hat{a}(\omega_2)\hat{a}_3^\dagger(\omega_1+\omega_2) + hc \qquad (44)$$

where

$$\Phi(\omega_1,\omega_2) = \operatorname{sinc}\left(\frac{L}{2}[k_1(\omega_1) + k_2(\omega_2) - k_3(\omega_1+\omega_2)]\right) \qquad (45)$$

For Type-I phase matching, wherein the incident photons are indistinguishable, we drop the subscripts and approximate to lowest order,

$$\Phi(\omega_1,\omega_2) = \operatorname{sinc}\left(\frac{k''L}{2}(\omega_1-\omega_0)(\omega_2-\omega_0)\right) \qquad (46)$$

where $k'' = (d^2k/d\omega^2)_{\omega_0}$. Time evolution over one pulse yields the 'output' operator

$$\hat{a}_3(\omega_3,out) = \hat{a}_3(\omega_3,0) + \xi \int d\omega_1 \Phi(\omega_1,\omega_3-\omega_1)\hat{a}(\omega_1)\hat{a}(\omega_3-\omega_1) \qquad (47)$$

where $\xi$ is proportional to $\xi'$ and

$$\Phi(\omega_1,\omega_3-\omega_1) = \operatorname{sinc}\left(\frac{k''L}{2}(\omega_1-\omega_0)(\omega_3-\omega_1-\omega_0)\right) \qquad (48)$$

With initial state $|\Psi\rangle = |vac\rangle_3 |squeezed\rangle_a$ we find the spectrum of the SFG field to be

$$\begin{aligned}
&\langle \hat{a}_3^\dagger(\omega_3,out)\hat{a}_3(\omega_3,out)\rangle_{QM} = \\
&\xi^2 \left\langle \int d\omega_1 \Phi^*(\omega_1,\omega_3-\omega_1)\hat{a}^\dagger(\omega_1)\hat{a}^\dagger(\omega_3-\omega_1) \int d\tilde{\omega}_1 \Phi(\tilde{\omega}_1,\omega_3-\tilde{\omega}_1)\hat{a}(\tilde{\omega}_1)\hat{a}(\omega_3-\tilde{\omega}_1)\right\rangle = \qquad (49)\\
&\xi^2 \int d\omega \int d\tilde{\omega}\, \Phi^*(\omega,\omega_3-\omega)\Phi(\tilde{\omega},\omega_3-\tilde{\omega})\langle \hat{a}^\dagger(\omega)\hat{a}^\dagger(\omega_3-\omega)\hat{a}(\tilde{\omega})\hat{a}(\omega_3-\tilde{\omega})\rangle_{QM}
\end{aligned}$$